\documentclass[pra,preprint,amssymb,showpacs]{revtex4}
\usepackage{amsmath}
\usepackage{dcolumn}
\usepackage{bm}
\usepackage{graphicx}
\def\be{\begin{equation}}
\def\ee{\end{equation}}
\def\bea{\begin{eqnarray}}
\def\eea{\end{eqnarray}}
\def\source#1#2#3#4{{\it #1}~{\bf #2}, #3 (#4)}
\def\Eq#1{Eq.~\ref{#1}}
\def\Eqs#1#2{Eqs. \ref{#1} and \ref{#2}}

\def\CA{{\cal A}}
\def\CM{{\cal M}}

\def\al{\alpha}
\def\om{\omega}
\def\irt{\frac{1}{\sqrt{3}}}
\def\half{{1 \over 2}}
\def\third{{1 \over 3}}
\def\GHZ#1{|\Psi_#1 \rangle}
\def\ket#1{| #1 \rangle}
\def\bra#1{\langle #1 |}

\def\BX{{\bf X}}

\begin{document}
\title{Mermin inequalities for perfect correlations in many-qutrit systems}
\author{Jay Lawrence} 
\affiliation{Department of Physics and Astronomy, Dartmouth
          College, Hanover, NH 03755, USA}
\affiliation{The James Franck Institute, University of Chicago, 
          Chicago, IL 60637}
\date{revised \today}
\bigskip
\begin{abstract}
The existence of GHZ contradictions in many-qutrit systems was a long-standing 
theoretical question until it's (affirmative) resolution in 2013. 
To enable experimental tests, we derive Mermin inequalities from concurrent 
observable sets identified in those proofs.  These employ a weighted sum of 
observables, called $\CM$, in which every term has the chosen GHZ state as an 
eigenstate with eigenvalue unity.  The quantum prediction for $\CM$ is then just 
the number of concurrent observables, and this grows asymptotically as $2^N/3$ as
the number of qutrits $N \rightarrow \infty$.  The maximum classical value falls short 
for every $N \geq 3$, so that the quantum to classical ratio (starting at 1.5 when 
$N=3$), diverges exponentially ($\sim 1.064^N$) as $N \rightarrow \infty$, where 
the system is in a Schr\"{o}dinger cat-like superposition of three macroscopically 
distinct states.

\end{abstract}
\maketitle
\section{Introduction}

Bell's inequality \citep{Bell64} shows that no local hidden variable theory (HV) can 
duplicate the quantum predictions for the correlations of two distant spin-1/2 particles 
(in Bohm's model \cite{Bohm51} of the original EPR scenario \cite{EPR35}).  Specifically,
the maximum quantum value of a certain correlation operator exceeds the maximum 
value allowed by HVs, with both quantum and HV predictions being probabilistic.  It was
not known for another quarter century (1964-1989) whether a stronger theorem existed, 
allowing for a definite (non-probabilistic) quantum prediction, until Greenberger, Horne, 
and Zeilinger (GHZ) \cite{GHZ89} found one for a  system of three spin-1/2 particles \cite{GHZ89footnote}.
Here, the product of three spin projections measured at distant points is predicted to 
take a single definite value, despite the randomness of the local measured values.  
The definiteness of the quantum prediction elevates HVs to the status of EPR elements 
of reality, since knowledge of local observables at two distant points allows prediction 
``with certainty'' of that at the third point.  On a practical level, this definiteness is 
essential in quantum information protocols such as quantum error correction 
\cite{Shor96} and quantum secret sharing \cite{BBH99}. 

In 1990, Mermin \cite{Mermin90} generalized the GHZ proof and supplied a Bell 
inequality for all $N \geq 3$ based on the perfect correlations predicted by quantum 
mechanics.   This was done to enable future experimental tests of GHZ contradictions 
by accounting for inevitable uncertainty in actual measurements, despite their absence 
in principle.   Experimental tests have indeed made use of such inequalities (now called
Mermin inequalities) to demonstrate GHZ contradictions with a probability of many
standard deviations.  The first such test \cite{Pan00} came a decade later;  a recent 
test \cite{Su17} describes the current state of the art. 

Extensions of Mermin's work within qubit sytems include GHZ contradictions based
on stabilizer groups of particular error-correcting codes \cite{DiVP97}, and
Mermin-like inequalities based on stabilizers for all graph states of $N \leq 6$  
\cite{Cabello08}.  In the latter work (2008),  Cabello et. al. defined a Mermin 
inequality as a Bell inequality for which (I) the Bell operator is a sum of stabilizing 
operators that represent the perfect correlations in their simultaneous eigenstate,
and (II) the ratio of quantum to maximum classical value is a maximum 
for that state.  We shall propose a modest extension below.

Equally interesting for us are the extensions to higher dimensions ($d$), which differ 
for even and odd cases.   Extensions to higher {\it even} dimensions include GHZ 
contradictions and Kochen-Specker identities \cite{CMP02,Mermin90b} for {\it odd}
$N>d$ (2002), then GHZ contradictions for odd $N<d$ \cite{LLK06} (2006), and 
more recently (2013), GHZ contradictions and corresponding Mermin inequalities 
\cite{Tang13} for systems of {\it all} $N \geq 4$ (with even $d$),  using GHZ-type 
graph states.

Regarding {\it odd dimensions}, Bell inequalities have been derived for systems 
of three \cite{Acin04} or more \cite{Alsina16} qutrits, as well as for higher odd $d$ 
\cite{Son06}. However, these are not Mermin inequalities; their quantum predictions 
are not definite, so they do not establish an underlying GHZ contradiction.  In fact, 
prior to (2013), it was not known whether a GHZ contradiction existed for any odd 
$d$.  It is now known that they do exist \cite{Ryu13, JL14}, and their discovery 
led to the completion of the program to establish GHZ contradictions (theoretically)
for all $N \geq 3$ for every $d \geq 2$ \cite{JL14}.  However, it is also known that 
these odd-$d$ contradictions cannot be based on stabilizer sets \cite{Howard13}, 
as is typical in even dimensions -- a conclusion drawn from studies of the discrete 
Wigner function for odd dimensions \cite{Gross,Veitch}.

The newly discovered GHZ contradictions in odd-$d$ systems \cite{Ryu13, JL14} are
based on concurrent observables \cite{LLK06} - observables that are not compatible 
but have a common eigenstate.  These are not the stabilizers usually associated 
with graph states, both because they lack compatibility and because local 
measurement bases are not exclusively those of the generalized Pauli operators.  
However, these aspects do not comprise experimental testing, for which the 
essential distinction between Bell and Mermin inequalities is the definiteness of 
the quantum predictions.  Thus, it seems appropriate to broaden the definition of
a Mermin inequality by deleting the word ``stabilizing'' from statement I (paragraph 3) 
above.

The purpose of this article is to construct Mermin inequalities, in this broader sense,
for systems of $N \geq 3$ qutrits, from sets of concurrent observables that share a 
GHZ eigenstate, violations of which would establish the perfect correlations of GHZ 
contradictions.  Entangled multiple-qutrit states are now being investigated 
experimentally \cite{Anton16}, and such inequalities will enable experimental tests 
of GHZ contradictions.  In the next section we present results for all $N \geq 4$; the
exceptional case of $N=3$ is presented in Sec. III, and in Sec. IV we discuss 
conclusions and open questions. 

\section{Mermin inequalities for $N \geq 4$}

It will be useful to consider three choices of GHZ state that differ by relative phases 
of components,
\be
   \GHZ{k} =  \irt \big( \ket{00...0} + \al^k \ket{11...1} + \al^{2k} \ket{22...2} \big),
   \hskip1.0truecm          (k = 0,1,2),
\label{GHZstates}
\ee
where $\al = \exp(2\pi i/9)$.   Envisioning the qutrits as spin-1 particles, Fig. 1a 
illustrates that $\GHZ{1}$ is obtained from $\GHZ{0}$ (and $\GHZ{2}$ from $\GHZ{1}$) 
by rotations of $2\pi/9 = 40^o$.   Such  ``rotations'' refer to any combination of 
individual qutrit rotations about their respective $\hat{z}$ axes, by increments 
adding up to $2\pi/9$.   A defining symmetry of GHZ states is that the rotated state 
is independent of the distribution of these increments among qutrits \cite{JL14}. 

\begin{figure}
\includegraphics[scale=.70]{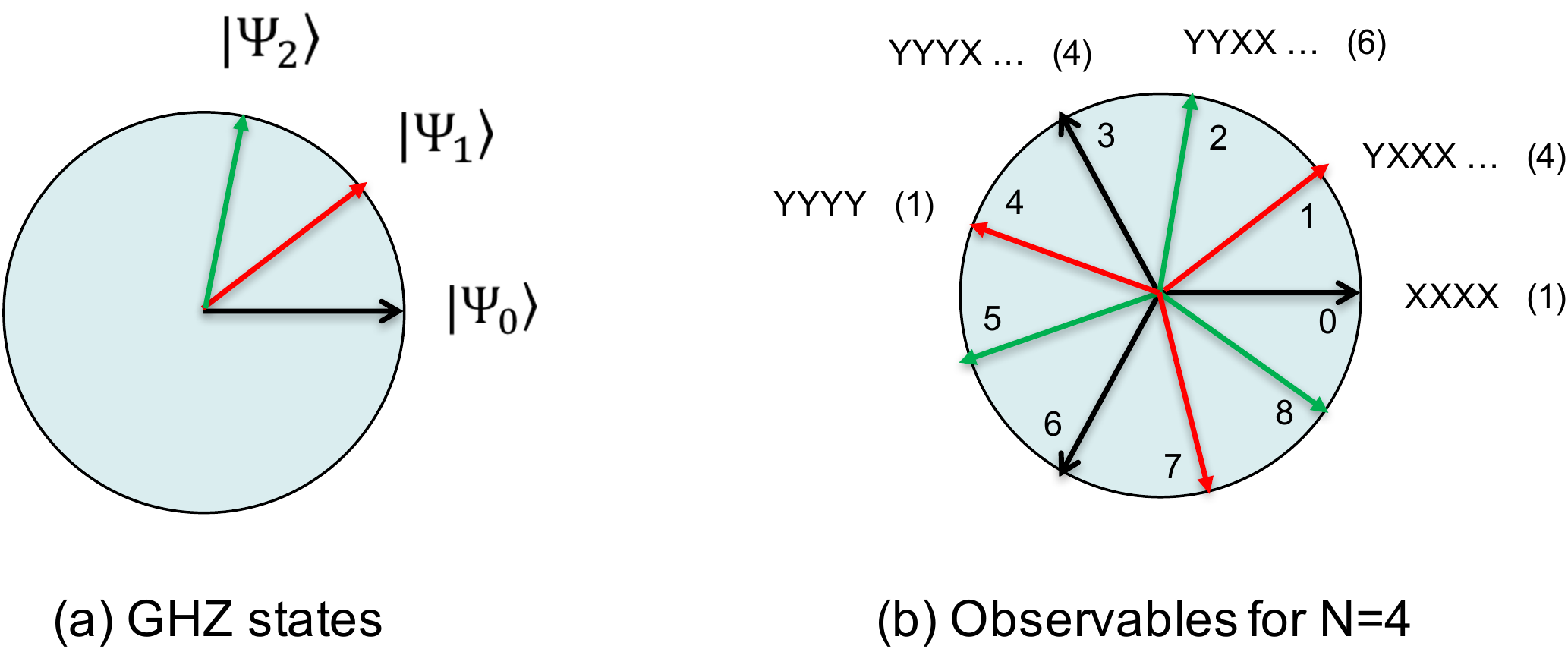}
\caption{\label{fig2} (a) GHZ states (\Eq{GHZstates}), and (b) tensor product 
observables for N=4.   Parentheses denote the number of permutations.  Black
arrows define the concurrent subset (of five observables) whose joint eigenstate 
is $\GHZ{0}$.} 
\end{figure}

The corresponding observable sets of which $\GHZ{k}$ are joint eigenstates are 
also related by compound rotations.  The starting point is the basic observable, 
\be
   \BX \equiv X^{\otimes N} = X_1...X_N,
\label{BigX}
\ee
where each factor $X_i$ is the qutrit Pauli matrix 
($X = \sum_{n=0}^2 \ket{n+1} \bra{n}$) that defines the first local measurement 
basis.   The second local basis ($Y_i$) is defined by a $2\pi/9$ rotation of $X_i$,
\be
   Y \equiv Z^{1/3} X Z^{-1/3} =  \sum_{n=_0}^2 \ket{n+1} \al^{(1 - 3\delta_{n,2})}\bra{n}, 
       \hskip0.7truecm   \hbox{where}  \hskip0.4truecm 
    Z = \sum_{n=0}^2 \ket{n} \om^n \bra{n}. 
\label{Ymatrix}
\ee
Compound rotations of the operator $\BX$ (\Eq{BigX}) through $2 \pi k/9$ generate 
tensor products in which $k$ factors of $Y$ are distributed in all possible ways,
$\big({k \atop N}\big)$, among $N-k$ factors of $X$.  These operators appear at  
points $k = 0,..., 8$ in Fig. 1b. 

Clearly, $\GHZ{0}$ is an eigenstate of $\BX$ with eigenvalue 1.  Rotational covariance 
of operators and states \cite{JL14} means that $\GHZ{1}$ is an eigenstate of all 
operators at the point 1, and $\GHZ{2}$ of all operators at point 2 - in all cases with 
eigenvalue 1.  Points 3 and beyond are then governed by the periodicity property 
\cite{JL14}:  Any rotation of an operator through $2\pi/3$ ($eg$, from 0 to 3) 
preserves its eigenstates, but multiplies its eigenvalues by $\om$.  Therefore, 
$\GHZ{0}$ is a joint eigenstate of operators at points 0, 3, and 6 (black arrows in 
Fig. 1);  $\GHZ{1}$ of operators at 1, 4, and 7 (red arrows); and $\GHZ{2}$ of operators 
at 2, 5, and 8 (green arrows).  In each case, the eigenvalues are 1, $\om$, and $\om^2$, 
respectively.  We shall refer to equilateral triangles (0,1,2) defined by each set of arrows, 
with concurrent operators at its vertices.   (Not all vertices are occupied when $N<8$.)

The case of $N=3$ is special because a third local measurement basis is required 
for GHZ contradictions.   Hence we defer that case and proceed here with $N =  4$, 
which is the simplest case.   Choosing the state $\GHZ{1}$, red arrows in Fig. 1b 
identify the concurrent subset of five observables -- the four cyclic permutations of 
$YXXX$, each with eigenvalue 1, and $YYYY$, with eigenvalue $\om$.  These 
observables produce a GHZ contradiction \cite{JL14}.   For the corresponding 
inequality, we define the Mermin operator,
\be
   \CM_1 = (YXXX + \hbox{permutations}) + \om^2 YYYY,
\label{Mwith4}
\ee
of which $\GHZ{1}$ is clearly an eigenstate with eigenvalue $\CM_Q = 5$.  This 
``quantum value'' is to be compared with the maximum HV value.  The {\it general} HV 
value, which we call $v(\CM_1)$, depends on the values (1, $\om$, or $\om^2$) 
assigned to each of the local factors [thence called $v(X_i)$  and $v(Y_i)$], which 
must be the same in each of the five tensor products.   It is easy to see that 
$|v(\CM_1)|$ depends only on the local ratios, 
\be
   R_i = v(Y_i)/v(X_i),
\label{ratios}
\ee
where, hiding an irrelevant overall phase factor, $v(\BX)$, it is simply
\be
   |v(\CM_0)| = |R_1 + R_2 + R_3 + R_4 + \om^2 R_1R_2R_3R_4|.
\label{ratio}
\ee
It is easy to verify by explicit calculations that the maximum value, $\CM_{HVM}$, is
obtained with either of two HV models: (i) uniform $R_i$ ($eg$, $R_i = 1$), and (ii)
a single departure from uniformity ($eg$,  $R_1 =  \om$ and all others $= 1$).  This  
maximum value is 
\be
   \CM_{HVM} = |4 + \om^2| = \sqrt{13} \approx 3.61,
\label{HVMax4}
\ee
so that the ratio of quantum to maximum HV values is
\be
   \CA = \CM_Q/\CM_{HVM} = 5/\sqrt{13} \approx 1.39.
\label{ratioA4}
\ee
Clearly the alternative choice, $\GHZ{0}$, together with operators at points 0 and 3,
would result in the same values of $\CM_Q$ and $\CM_{HVM}$, while the choice 
$\GHZ{2}$, with operators only at point 2, would result in $\CM_Q = \CM_{HVM} = 5$, 
showing no GHZ contradiction.  Therefore, $\CM_0$ and $\CM_1$ are equally valid
Mermin operators, according to the definition. 

For arbitrary $N  > 4$, we pick a state $\GHZ{k}$ and identify its candidate Mermin 
operator $\CM_k$ as the sum all concurrent operators at the vertices of the 
corresponging triangle, with weighting factors 1, $\om^2$, and $\om$ assigned to first, 
second, and third vertices traversed in counterclockwise order.  As above, the results 
($\CM_Q$ and $\CM_{HVM}$) depend on the choice of $k$, and Table I shows those 
choices which maximize the ratio $\CA$ and produce Mermin operators.   For even 
$N$, by symmetry, there are two such choices; for odd $N$, only one.  The 
corresponding quantum eigenvalues are given by
\be
   M_Q = \third (2^N -1)   \hskip.5truecm  (\hbox{even}~N);   \hskip1.2truecm
   M_Q = \third (2^N -2)   \hskip.5truecm  (\hbox{odd}~N);
\label{MsubQ}
\ee
equal to the total number of concurrent observables on the $k$th triangle. 
\begin{table}[tbp]
\caption{Quantum and maximum HV values of the $N$-qutrit Mermin operator, and
their ratio $\CA$, as functions of $N$.  Listed values of $k$ are those which 
maximize $\CA$.}
\begin{equation*}
\begin{tabular}{|c|c|ccc|}
\hline
  \  $N$  \  &  \  $k$  \  & \ $\CM_Q$  \  &  \  $\CM_{HVM}$  \  &  \  $\CA$  \\   \hline
   \   4   \  &   \  0,1 \ &  \ 5   \ &  \ $\sqrt{13}$ \ &  \ 1.39  \         \\
   \   5   \  &   \  1    \ &  \ 10  \ &  \ 7                 \ &  \ 1.43   \        \\
   \  6    \  &   \  1,2 \ &  \ 21  \ &  \ $3\sqrt{19}$ \  & \ 1.61  \       \\
   \   7   \  &   \   2    \ &  \ 42  \ &  \ 24                  \  & \ 1.75   \        \\
   \   8   \  &   \  2,0  \ &  \ 85  \ &  \ $\sqrt{2269}$ \ & \ 1.78   \    \\
   \   9   \  &   \   0    \ &   \  170 \ & \ $\sqrt{6892}$ \ & \ 2.05   \    \\
  \ \    10  \ \  &   \ \   0,1 \ \  & \   341  \ & \ $\sqrt{29,791}$ \  & \  1.98  \         \\
  \ \    11  \ \  &    \ \   1    \ \  &  \  682   \ &   \   308                 \  & \  2.21   \        \\
  \ \    12  \ \  &   \ \  1,2  \ \  &  \ 1365   \ &   \ $\sqrt{385,947}$  \ & \ 2.20   \        \\
  \ \    13  \ \  &    \ \  2     \ \  &   \  2730    &    \  1131                    \ & \  2.41   \       \\
\hline
\end{tabular}
\end{equation*}
\end{table}

The contrasting $\CM_{HVM}$ values are maxima of $|v(\CM_k)|$, given $k$ on the
Table, over the local ratios $R_i$ (\Eq{ratios}).  In the following six paragraphs, we 
show how these are determined, including the proof of the following --

\noindent {\bf Theorem:}  Maximum values of $|v(\CM_k)|$, for $k$ values listed on 
Table I, are attained with uniform $R_i$ in all cases except $N = 5$, 7, and 9, where 
the simplest nonuniform model ($R_1 = \om$, $R_2 ... R_N = 1$) narrowly prevails.   

\noindent {\bf Proof:}   A closed-form expression for candidate Mermin operators 
(valid for $k =0,$ 1, or 2) is \cite{JMK16}
\be
   \CM_k = \third \bigg[ ( X + \al^2 Y)^N + \om^{2k} (X + \om \al^2 Y)^N + 
                                        \om^k (X + \om^2 \al^2 Y)^N \bigg].
\label{Mermingen}
\ee
To verify, one can easily see that certain powers of $Y$ arising in the binomial 
expansions cancel out because $1+\om+\om^2 =0$.  With $k=0$, for example, this 
cancellation leaves only the powers 0, 3, 6, ..., exactly those terms residing on the
vertices of the $k = 0$ triangle in Fig. 1b; and similarly for $k = 1$ and 2.  One can 
also verify that the relative vertex weighting factors are (1, $\om^2$, and $\om$), as 
required, once the higher powers of $\al$ have been reduced ($eg$, 
$\al^4 = \om \al$, $etc$.).   (Note that $\CM_k$ appears with overall multiplying 
factor $\al^{2k}$.)

The main utility of \ref{Mermingen} is to make the $R_i$-dependence explicit in
\be
   |v(\CM_k)| = \third  \left| \prod_{i=1}^N  (1 + \al^2 R_i)^N  +
                  \om^{2k} \prod_{i=1}^N   (1 + \om \al^2 R_i)^N   +     
                  \om^k  \prod_{i=1}^N  (1 + \om^2 \al^2 R_i)^N \right|.
\label{FMHV}
\ee
If the $R_i$ are uniform, then 
\bea
 &  |v(\CM_k)_{unif}| = \third \left| (1 + \al^2)^N + \om^{2k} (1 + \om \al^2)^N + 
                                                \om^k (1 + \om^2 \al^2)^N \right|  \nonumber    \\
                     & = \third \left| B^N \exp(2 \pi N i/9) + \om^{2k} C^N \exp(- 4 \pi N i/9) +
                                                \om^k A^N \exp(- \pi N i/9)  \right|,
\label{FMHVunif}
\eea  
where the individual magnitudes are labeled so that $A > B > C$:
\bea
  &  A = |1 + \om^2 \al^2| = 2 \cos {\pi \over 9} \approx 1.8794,  \nonumber   \\
  &  B = |1 + \al^2| = 2 \cos {2 \pi \over 9} \approx 1.5321,           \nonumber   \\
  &  C = |1 + \om \al^2| = 2 \cos {4 \pi \over 9} \approx 0.3473,  
 \label{magnitudes}
\eea
and the identity $1 + e^{i \theta} = \cos {\theta \over 2} e^{i \theta/2}$ was used.   If a 
single $R_i$ differs from the rest ($eg$, $R_1 = \om$, others unity), then the effects 
on \Eq{FMHVunif} are to permute the coefficients, $B_1 \rightarrow C_1 \rightarrow 
A_1 \rightarrow B_1$ and to rotate each vector in the complex plane:
\bea
   &  |v(\CM_k)_{R_1=\om}| = \third \mid B^{N-1}C \exp(2 \pi N i/9) \exp(-2 \pi i/3)  
         \nonumber  \\  + & \om^{2k} C^{N-1}A \exp(-4 \pi N i/9) \exp(\pi i/3)      
          + \om^k A^{N-1}B \exp(- \pi N i/9) \exp(\pi i/3) \mid ,
\label{FMHVoneout}
\eea
where the rotation angles [($-2 \pi/3$), ($\pi/3$), and ($\pi/3$), respectively] are 
independent of which $R_i$ is chosen to be different.   
   
Thus, introducing the nonuniformity decreases the two largest terms while 
increasing only the smallest.  This can produce a net gain in $|v(\CM_k)|$ only 
if the rotations bring the two largest terms into closer alignment.  This unlikely 
scenario is actually realized in the few cases, $N=5$, 7, and 9.  
 
To demonstrate, first consider odd $N$.  With $k$ values listed on Table I, it is 
easy to see that the three vectors comprising $v(\CM_k)$ in either \ref{FMHVunif} or 
\ref{FMHVoneout} are collinear for every odd $N \geq 5$.  In \ref{FMHVunif}, the $A$ 
term is aligned opposite to the $B$ and $C$ terms, so that
\be
    |v(\CM_k)_{unif}| = \third \bigg( A^N - B^N - C^N \bigg).
\label{Voddunif}
\ee
In \ref{FMHVoneout}, the $C$-like term is aligned opposite to the others, and so
\be
   |v(\CM_k)_{R_1=\om}| = \third \bigg( A^{N-1}B + B^{N-1}C - C^{N-1}A \bigg).
\label{Voddoneout}
\ee
The difference,  (\ref{Voddunif} - \ref{Voddoneout}), is an increasing function of $N$ with a 
zero at $N_o \approx 9.26$, so that $|v(\CM_k)_{unif}|$ is the larger for all odd $N \geq 11$, 
while $|v(\CM_k)_{R_1=\om}|$ is the larger for 5, 7, and 9.  We still have to rule out more
complex HV models -- this is done below.

Now consider even $N$:  Again with $k$ values listed on Table I, one can easily see 
that the three vectors in \Eq{FMHVunif} (uniform $R_i$) are minimally aligned in the 
complex plane (angular separations are $2 \pi/3$).  Nonuniformity (\ref{FMHVoneout}) 
shrinks the two longer vectors as above, while the induced rotations improve their 
alignment somewhat (to the smaller of the angular separations $\pi/3$, $\pi/3$, and 
$4 \pi/3$), but not enough to provide a net gain in $|v(\CM_k)|$:  Keeping the two 
dominant terms in each of \Eqs{FMHVunif}{FMHVoneout}, whose angular separations 
are $2 \pi/3$ and $\pi/3$, respectively, it is easy to show formally that 
$|v(\CM_k)_{unif}| - |v(\CM_k)_{R_1=\om}|$ is positive for all $N \geq 6$.  For $N=4$, 
the exact calculations described above show that both HV models realize the 
maximum value.  Hence $|v(\CM_k)_{unif}|$ provides the maximum for all even $N$.

To rule out further HV models for all even and odd $N \geq 4$:  First consider the  
alternate single departure, ($R_1=\om^2$, others unity).  The largest term is reduced 
sharply (by $C/A$), the next largest is increased slightly (by $A/B$), while the relative
alignment of these two remains unchanged.  So this model is 
ruled out trivially.  Multiple departures from uniformity may be viewed as sequences 
of single departures in which every step has the following properties:  (i) Either it 
reduces the two longest vectors, or it reduces the longest by more than it increases 
the next-longest, and (ii) beyond the first step (which results in \Eq{FMHVoneout}), it 
reproduces angular separations already seen in \Eq{FMHVunif} or \ref{FMHVoneout}.  
Thus it cannot increase $|v(\CM_k)|$ beyond the larger of $|v(\CM_k)_{unif}|$ and 
$|v(\CM_k)_{R_1=\om}|$.

This concludes the proof of the theorem stated above.  To evaluate $\CM_{HVM}$, 
one may simply use \Eq{Voddunif} or \ref{Voddoneout} for odd $N$; for even $N$
use \ref{FMHVunif}, knowing the the angular separations are $2\pi/3$ for the listed
$k$-values.   It is also instructive for smaller $N$ to write the Mermin operator 
directly from Fig. 1b and evaluate at $\{R_i\}$ determined by the theorem.  Table I 
lists the exact maxima, $\CM_{HVM}$, which are all integers or square roots thereof, 
along with rounded values of $\CA$.

The asymptotic form of $\CM_{HVM}$ at large $N$ is given by the dominant term in 
\ref{FMHVunif}, namely
\be
   \lim_{N \rightarrow \infty} \CM_{HVM} = \third A^N \approx \third 1.879^N,
\label{asymptote}
\ee
so that the quantum to classical ratio (\Eq{MsubQ} to \ref{asymptote}) diverges as 
$1.064^N$.  This exponential divergence is slow compared with Mermin's ($2^{N/2}$) 
for qubit systems \cite{Mermin90};  nevertheless it represents a superposition of three macroscopically distinct states.

\section{The case of N = 3}

This may be the most interesting case experimentally.  It is singled out here because 
its GHZ contradictions require three local measurement bases \cite{Ryu14,JL14}.  So,
while we consider the same three GHZ states (\ref{GHZstates}), the concurrent
operator sets must now incorporate a third local basis, a natural choice being given 
by rotation of individual $X$ factors through $4 \pi/9$: 
\be
   W \equiv Z^{2/3} X Z^{-2/3} = \sum_{n=_0}^2 \ket{n+1} \al^{(2 - 6 \delta_{n,0})}\bra{n}
\label{Wmatrix}
\ee
(compare \ref{Ymatrix}).   The observables generated by rotations of $XXX$ now
include all combinations of $X$, $Y$, and $W$ factors, and are classified in Fig. 2 
according to total rotation angles, $2\pi k/9$.  Again these fall into three concurrent 
subsets, each associated with an equilateral triangle and its own joint eigenstate 
in Fig. 1a.  In this case all three Mermin operators $\CM_k$ produce the same 
outcome:  $\CM_Q = 9$ and $\CM_{HVM} = 6$.  Let us demonstrate with the 
simplest example:
\be
    \CM_0 = \bigg[ XXX + \om^2 (YYY + XYW + XWY + YXW + WXY + YWX + WYX) + 
                \om WWW \bigg],
\label{Mthree0}
\ee
with weight factors (1,$\om^2,\om$) applied as required.  Recall that the HV 
%
\begin{figure}
\includegraphics[scale=.70]{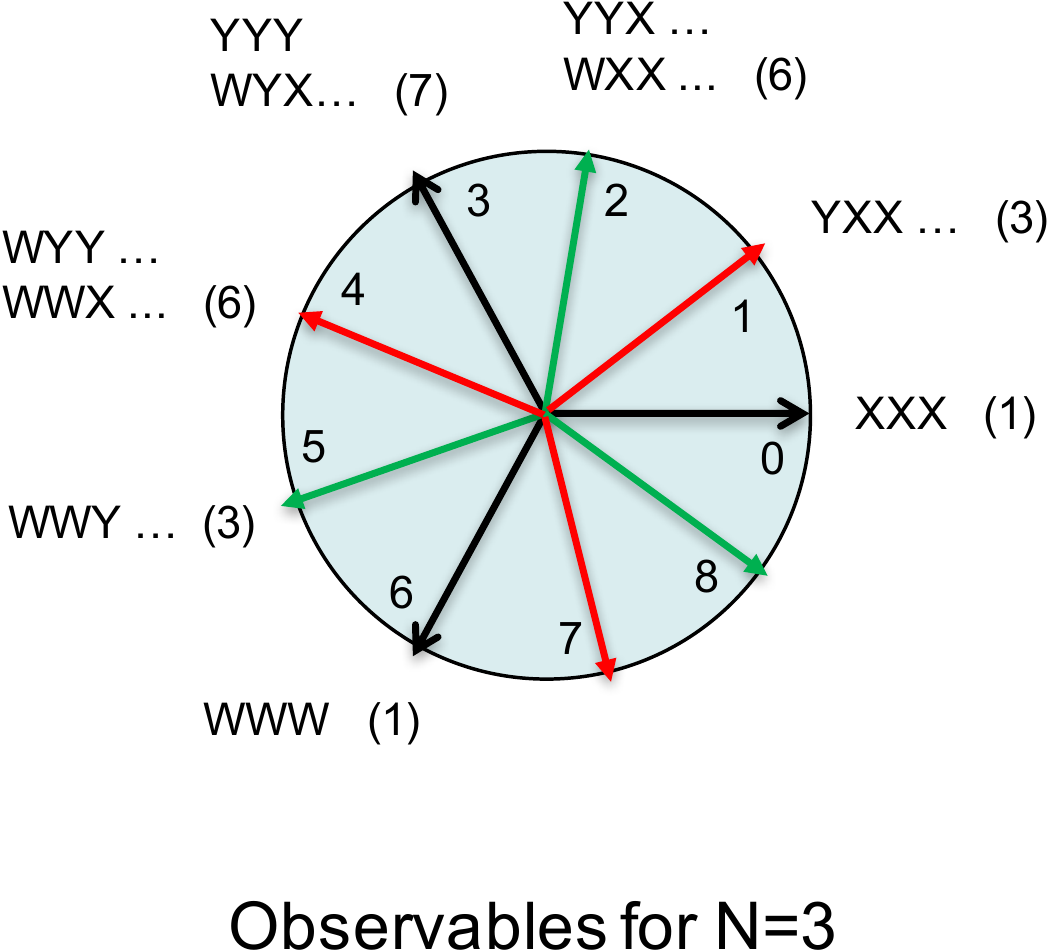}
\caption{\label{fig2} Tensor product observables for N=3 form three concurrent
subsets.  Each produces a Mermin inequality with $\CM_Q = 9$ and 
$\CM_{HVM} = 6$.} 
\end{figure}
magnitude depends only on ratios, defined here as $R_i = v(X_i)/v(Y_i)$ and 
$S_i = v(W_i)/v(Y_i)$.  So, hiding an overall irrelevant phase factor $v(YYY)$,
\be
   |v(\CM_0)| = | RRR + \om^2 (111 + R1S + R1S + 1RS + SR1 + 1SR + S1R)
                                           + \om SSS |,
\label{Vthree0}
\ee
where the subscripts of $R_i$ and $S_i$ are implied by their positions, $eg$, 
$R1S = R_1 S_3$.   Now suppose the ratios are uniform, and $R_i = S_i = 1$.  Then,
$|v(\CM_0)| = |1 + 7 \om^2 + \om| = 6$.  To increase this value, one would require an 
HV assignment that brought the first and/or last term into equality with the seven 
other terms, without losing an equal number (or more) of those terms.  It is easy to 
see that there is no such assignment.

Finally, it is interesting to note that $|v(\CM_1)|$ and $|v(\CM_2)|$ yield the same 
maximum, but both require nonuniform HV assignments, eg., ($R_1 = S_1 = \om$ 
with all others unity) for the former, and, for the latter, ($S_1 = \om$ with all others 
unity).


\section{Conclusions and open questions}

We have presented Mermin operators and associated inequalities for systems 
of $N \geq 3$ qutrits.   The exceptional case of $N=3$ requires three local 
measurement bases; all other cases require two.  The eigenvalue of the Mermin 
operator (the definite quantum prediction of its measured value), is given by 
\Eq{MsubQ} and diverges as $2^N/3$ for large $N$.  The maximum HV values 
are illustrated in Table I and reflect optimal HV assignments derived in Sec. II. 
These diverge as 1.879$^N$.   The ratio of quantum to maximum HV values 
diverges as $1.064^N$.   

Ironically, the  structure behind the inequalities derived here forms a close parallel 
with Mermin's, despite the compatibility of his observable sets  as compared with the 
mere concurrence of those used here.   This is because his Pauli tensor products 
and their eigenstates are related by the same rotational covariance that forms the 
basis of the treatment given here.  It is a simple exercise to write down two 
alternative compatible Pauli subsets (one of which is Mermin's), and their 
corresponding joint GHZ eigenstates, on diagrams analogous to Fig. 1, in which the 
basic angular interval is $\pi/2$ rather than $2\pi/9$.  Moreover, Mermin's derivation 
of HV maxima is based on a formula like our \Eq{Mermingen}, in particular
\be
   \CM_k^{d=2} = \half \bigg[ ( X + i Y)^N + (-1)^k (X - i Y)^N \bigg],
\label{Mermingen2}
\ee
obtained by replacing $\om \rightarrow -1$ and  $\al \rightarrow \exp (i \pi/4)$.   
The two choices $k=0$ and 1 produce identical $\CM_Q$ and $\CM_{HVM}$ 
values, resulting in $\CA = 2^{N/2}$ (even $N$), and $2^{(N-1)/2}$ (odd $N$). 

A comparison of \Eqs{Mermingen}{Mermingen2} suggests why our exponential 
growth $\CA \rightarrow 1.064^N$ is less dramatic than Mermin's.  The maximum
length of any factor in the HV expression for qutrits (\ref{FMHV}) is $A =
|1 + \om^2\al^2|^{1/2} = 2\cos \pi/9 \approx 1.8794$, no matter how HV values are
assigned.  The analogous length factor in the qubit case is $\sqrt{2}$.  This 
corresponds to the different angular resolutions of vector factors in 
\Eqs{Mermingen}{Mermingen2}, showing minimum angles in the complex plane of 
$\pi/9$ {\it vs} $\pi/4$.  These differences reflect the greater freedom of qutrit HVs 
over qubit HVs in aiming for the quantum results.

The above comparison raises the question whether Mermin inequalities exist for
systems of higher odd dimensions $d$, where compatible observables do not 
produce GHZ contradictions.   It seems plausibible that a similar construction would
succeed for any higher prime $d$, although one would expect still weaker 
violations of local realism for the reason given above.  For higher composite 
dimension, a similar but more complex construction might succeed based on 
the smallest prime factor of $d$.


\begin{thebibliography}{99}
\bibitem{GHZ89footnote} GHZ \cite{GHZ89} actually derived their result with four 
   spins-1/2, mentioning in a footnote that it also holds with three.
\end{thebibliography}
\end{document}